\documentclass[12pt]{article}

\usepackage{graphicx}
\usepackage{amsfonts}

\def\beq#1{\begin{equation}\label{#1}}
\def\eeq{\end{equation}}

\newcommand{\bear}[1]{\begin{eqnarray}\label{#1}}
\newcommand{\ear}{\end{eqnarray}}

\catcode`\@=11 \@addtoreset{equation}{section}\catcode`\@=12

\newcommand{\R}{{\mathbb R}}

\newcommand{\p}{\partial}

\newcommand{\nn}{\nonumber}

\begin{document}

 \begin{center}

 \large \bf
  Dilatonic dyon-like black hole solutions in the model with two Abelian gauge fields

  \end{center}

 \vspace{0.3truecm}

  \begin{center}
   M.E. Abishev$^{1,3}$, K.A. Boshkayev$^{1}$, \\
   and V.D. Ivashchuk$^{2,3}$    \\

    \vspace{0.3truecm}

   $^{1}$  Institute of Experimental and Theoretical Physics, \\
    Al-Farabi Kazakh National University,  \\
   Al-Farabi avenue, 71, Almaty 050040, Kazakhstan, \\
   $^{2}$ Center for Gravitation and Fundamental Metrology, VNIIMS, \\
   Ozyornaya St., 46, Moscow 119361, Russia, \\
   $^{3}$ Institute of Gravitation and Cosmology, \\
   RUDN University, \\
   Miklukho-Maklaya St.,6, Moscow 117198,  Russia

 \end{center}

\begin{abstract}
Dilatonic black hole dyon-like solutions in the gravitational
$4d$ model with a scalar field,  two 2-forms, two dilatonic coupling constants
$\lambda_i \neq 0$, $i =1,2$, obeying $\lambda_1 \neq - \lambda_2$
and the sign parameter $\varepsilon = \pm 1$ for scalar field kinetic term
are considered. Here $\varepsilon = - 1$ corresponds to a ghost scalar field. 
These solutions are defined up to solutions of two master equations for two moduli functions, when
 $\lambda^2_i \neq 1/2$ for $\varepsilon = - 1$.
Some physical parameters of the solutions are obtained:
gravitational mass, scalar charge, Hawking temperature, black hole area entropy and
parametrized post-Newtonian  (PPN) parameters $\beta$ and $\gamma$.
The PPN parameters do not depend on the couplings $\lambda_i$ and
$\varepsilon$. A set of bounds on the gravitational mass
and scalar charge  are found by using
a certain conjecture on the parameters of solutions,
when $1 +2 \lambda_i^2 \varepsilon > 0$, $i =1,2$.

\end{abstract}

\pagebreak

\normalsize

%%%%%%%%%%%%%%%%%%%%%%%%%%%%%%%%%%%%%%%%%%%%%%%%%%%%%%%%%%%%%%%%
\section{Introduction}
%%%%%%%%%%%%%%%%%%%%%%%%%%%%%%%%%%%%%%%%%%%%%%%%%%%%%%%%%%%%%%%%

In this paper we extend our previous work \cite{ABDI} devoted to dilatonic dyon black hole solutions.
We note that at present there exists a certain interest in spherically symmetric solutions,
 e.g. black hole and black brane ones, related to Lie algebras and
Toda chains, see \cite{BronShikin}-\cite{GKO}
and the references therein. These solutions appear
in gravitational models with scalar fields and antisymmetric forms.

Here we consider a subclass of
dilatonic black hole solutions with  electric and magnetic
charges $Q_1$ and $Q_2$, respectively, in the $4d$ model with metric $g$,  scalar field $\varphi$,  two 2-forms  $F^{(1)}$ and $F^{(2)}$, corresponding to two dilatonic coupling constants $\lambda_1$ and $\lambda_2$, respectively.
All fields are defined on an oriented manifold ${\cal M }$. 
Here we  consider the dyon-like configuration for fields of 2-forms:
\beq{0.1}
 F^{(1)} = Q_1 e^{2 \lambda_1 \varphi} * \tau, \qquad F^{(2)} = Q_2 \tau,
 \eeq
where $\tau = {\rm vol}[S^2]$ is volume form on 2-dimensional sphere and $* = *[g]$ is the Hodge operator
corresponding to the oriented manifold ${\cal M }$  with the metric $g$. 
We call this noncomposite configuration a  dyon-like one 
in order to distinguish it from the true dyon configuration which is essentially composite and
may be chosen in our case either as: (i) $F^{(1)} = Q_1 e^{2 \lambda_1 \varphi} * \tau + Q_2 \tau$, $F^{(2)} =0$,
or (ii) $F^{(1)} =0$, $F^{(2)} = Q_1 e^{2 \lambda_2 \varphi} * \tau + Q_2 \tau$. From a physical point of view 
the ansatz (\ref{0.1}) means that we deal here with a charged black hole, which has two color charges:
$Q_1$ and $Q_2$. The charge $Q_1$ is the electric one corresponding to the form $F^{(1)}$, while 
the charge $Q_2$ is the magnetic one corresponding to the form $F^{(2)}$. For coinciding dilatonic couplings 
$\lambda_1 = \lambda_2 = \lambda$ we get a trivial noncomposite generalization of 
dilatonic dyon black hole solutions in the model with one 2-form which was considered in ref. \cite{ABDI}, see also 
\cite{Lee,ChHsuL,GKLTT,PTW,FIMS,GKO} and references therein.

The dilatonic scalar field may be either an ordinary one or a phantom (or ghost) one.
The phantom field   appears in the action with a kinetic term of the
``wrong sign'', which implies the violation of the null energy condition $p  \geq - \rho
$. According to ref. \cite{ArkH}, at the quantum level, such fields could form a
``ghost condensate'', which may be responsible for modified
gravity laws in the infra-red limit. The observational data do not exclude this possibility \cite{Kom}.

Here we seek  relations for the physical parameters of dyonic-like black holes, e.g.
bounds on the gravitational mass $M$ and the scalar charge $Q_{\varphi}$. As in our previous work \cite{ABDI}
 this problem is  solved here up to a conjecture, which states a one-to-one (smooth) correspondence between
the pair  $(Q_1^2,Q_2^2)$, where $Q_1$ is the electric charge and $Q_2$ is the magnetic charge,
and the pair of positive parameters $(P_1,P_2)$, which appear in decomposition of
 moduli functions at large distances.
This conjecture is believed to be valid for all $\lambda_i \neq 0$
in the case of an ordinary scalar field and for $0< \lambda_i^2 < 1/2$ for the case of a phantom scalar field 
(in both cases the inequality $\lambda_1 \neq - \lambda_2$ is assumed).

%%%%%%%%%%%%%%%%%%%%%%%%%%%%%%%%%%%%%%%%%%%%%%%%%%%%%%%%%%%%%%%%
\section{Black hole dyon solutions}
%%%%%%%%%%%%%%%%%%%%%%%%%%%%%%%%%%%%%%%%%%%%%%%%%%%%%%%%%%%%%%%%

Let us consider a model governed by the action
\bear{i.1}
 S= \frac{1}{16 \pi G}  \int d^4 x \sqrt{|g|}\biggl\{ R[g] -
 \varepsilon g^{\mu \nu}\p_{\mu} \varphi  \p_{\nu} \varphi
 \qquad \qquad   \nn \\
 - \frac{1}{2} e^{2 \lambda_1 \varphi} F^{(1)}_{\mu \nu} F^{(1)\mu \nu }
 - \frac{1}{2} e^{2 \lambda_2 \varphi} F^{(2)}_{\mu \nu} F^{(2) \mu \nu}
 \biggr\},
\ear
where $g= g_{\mu \nu}(x)dx^{\mu} \otimes dx^{\nu}$ is  metric,
 $\varphi$ is the scalar field, $F^{(i)} = dA^{(i)}
 =  \frac{1}{2} F^{(i)}_{\mu \nu} dx^{\mu} \wedge dx^{\nu}$
is the $2$-form with $A^{(i)} = A^{(i)}_{\mu} dx^{\mu}$, $i =1,2$, $\varepsilon = \pm 1$,
$G$ is the gravitational constant,
 $\lambda_1, \lambda_2 \neq 0$ are  coupling constants obeying $\lambda_1 \neq - \lambda_2$ and
 $|g| =   |\det (g_{\mu \nu})|$. Here we also put $\lambda_i^2 \neq 1/2$, $i =1,2$,
 for $\varepsilon = - 1$. For $\lambda_1 = \lambda_2$ the Lagrangian  (\ref{i.1})
 appears in the gravitational model with a scalar field and Yang-Mills field 
 with a gauge group  of rank $2$ (say $SU(3)$) when an Abelian sector of the gauge field is considered. 

 We consider a family of dyonic-like black hole
solutions to the field equations corresponding to the action
(\ref{i.1}) which are defined on the manifold
\beq{i.2}
 {\cal M }  =    (2\mu, + \infty)  \times S^2 \times  \R,
\eeq
and have the following form
\bear{i.3}
 &&ds^2 = g_{\mu \nu} dx^{\mu} dx^{\nu}
 = H_1^{h_1} H_2^{h_2}
 \biggl\{ -  H_1^{-2 h_1} H_2^{-2 h_2}
 \left( 1 - \frac{2\mu}{R} \right)
 dt^2 \\ \nonumber
 && \qquad +  \frac{dR^2}{1 - \frac{2\mu}{R}} + R^2  d \Omega^2_{2}
  \biggr\},
 \\  \label{i.3a}
 &&\exp(\varphi)=
 H_1^{h_1 \lambda_1 \varepsilon} H_2^{- h_2 \lambda_2 \varepsilon },
 \\  \label{i.3be}
 &&F^{(1)}=
 \frac{Q_1}{R^2}   H_{1}^{-2} H_{2}^{- A_{12}} dt \wedge dR,
  \\  \label{i.3bm}
   &&F^{(2)}  =  Q_2 \tau.
\ear
Here  $Q_1$ and $Q_2$ are (colored) charges -- electric and magnetic, respectively,
 $\mu > 0$ is the extremality parameter,
 $d \Omega^2_{2} = d \theta^2 + \sin^2 \theta d \phi^2$
is the canonical metric on the unit sphere $S^2$
 ($0< \theta < \pi$, $0< \phi < 2 \pi$),
 $\tau = \sin \theta d \theta \wedge d \phi$
is the standard volume form on $S^2$,
\beq{i.16}
 h_i = K_i^{-1}, \qquad K_i = \frac{1}{2} +  \varepsilon \lambda_i^2,
\eeq
$i =1,2$, and
\beq{i.16a}
  A_{12} =  (1 - 2 \lambda_1 \lambda_2 \varepsilon) h_2.
\eeq
The functions $H_s > 0$ obey the equations
\beq{i3.1}
 R^2 \frac{d}{dR} \left( R^2
  \frac{\left(1 - \frac{2 \mu}{R}\right)}{H_s}
  \frac{d H_s}{dR} \right) = - K_s  Q_s^2
  \prod_{l = 1,2}  H_{l}^{- A_{s l}},
\eeq
with the following boundary conditions imposed:
\beq{i3.1a}
  H_s  \to H_{s0} > 0
\eeq
for $R \to 2\mu $, and
\beq{i3.1b}
  H_s    \to 1
\eeq
for $R \to +\infty$, $s = 1,2$.

In (\ref{i3.1}) we denote
\beq{i.18}
    \left(A_{ss'}\right)=
  \left( \begin{array}{*{6}{c}}
     2 &  A_{12}\\
     A_{21} & 2\\
\end{array}
\right) ,
\eeq
where $A_{12}$ is defined in (\ref{i.16a}) and
\beq{i.16b}
  A_{21} =  (1 - 2 \lambda_1 \lambda_2 \varepsilon) h_1.
\eeq

These solutions  may be obtained just by using  general formulas for non-extremal (intersecting)  black brane solutions from \cite{IMp1,IMp2,IMp3} (for a review see \cite{IMtop}).
The composite analogs of the solutions  with one 2-form and $\lambda_1 = \lambda_2$
 were presented  in ref. \cite{ABDI}.

The first boundary condition (\ref{i3.1a}) guarantees (up to a possible additional requirement on the analyticity 
of $H_s(R)$ in the vicinity of $R= 2 \mu$)
the existence of a (regular) horizon at  $R = 2 \mu$ for the metric (\ref{i.3}).
The second condition (\ref{i3.1b}) ensures  asymptotical (for $R \to
+\infty$) flatness of the metric.

{\bf Remark 1.} {\em It should be noted that the main motivation for considering this and more general $4D$ models 
governed by the Lagrangian  density ${\cal  L}$:
 \begin{equation}
  {\cal  L}/ \sqrt{|g|} =  
   R[g] -   h_{ab} g^{\mu \nu}\p_{\mu} \varphi^a  \p_{\nu} \varphi^b
    - \frac{1}{2} \sum_{i =1}^{m} \exp(2 \lambda_{i a}  \varphi^{a})F^{(i)}_{\mu \nu} F^{(i)\mu \nu},
    \label{i.16r}  
  \end{equation}
where $\varphi =(\varphi^a)$ is a set of $l$ scalar fields,  $F^{(i)} = d A^{(i)}$
are 2 forms and  $\lambda_{i} = (\lambda_{i a})$ are dilatonic coupling vectors,
$i =1, \dots, m$, is coming from dimensional reduction of supergravity models;
in this case the matrix  $(h_{ab})$ is positive definite.  For example, one may consider 
a part of bosonic sector of  dimensionally reduced  $11d$ supergravity \cite{LP} with $l$ dilatonic scalar fields and $m$ $2$-forms (either  originating  from 11d metric  or  coming from $4$-form) activated; Chern-Simons terms  vanish in this case.  Certain uplifts (to higher dimensions) of 4d black hole solutions corresponding to  
(\ref{i.16r}) may lead us to black brane solutions in dimensions $D >4$, 
e.g. to dyonic ones; see \cite{LP,DLP,IMp2,CGLO,GO} and the
references therein. 
The dimensional reduction from the 12-dimensional model from ref. \cite{KKLP} with phantom scalar field and two forms of rank $4$ and $5$ will lead us to the Lagrangian  density
(\ref{i.16r}) with the matrix $(h_{ab})$ of pseudo-Euclidean signature.} 

Equations (\ref{i3.1}) may be rewritten in the following form:
\beq{i2.1}
  \frac{d}{dz} \left[
  \left(1 - z \right) \frac{d y^s}{dz} \right] =
          - K_s  q_s^2 \exp(- \sum_{l =1,2} A_{sl} y^l ),
\eeq
 $s = 1,2$. Here and in the following we use the following notations:
 $y^s= \ln H_s$,  $z = 2 \mu/R$, $q_s = Q_s/(2\mu)$ and  $K_s = h_s^{-1}$
 for $s = 1,2$, respectively.
We are seeking solutions to equations (\ref{i2.1}) for  $z \in (0,1)$
obeying
 \bear{i2.1b}
   y^s(0) = 0, \\ \label{i2.1c}
   y^s(1) = y^{s}_0,
    \ear
where $y^{s}_0 = \ln H_{s0}$ are finite (real) numbers, $s = 1,2$.
Here $z=0$ (or, more precisely $z=+ 0$) corresponds to infinity ($R = + \infty$), while
$z=1$ (or, more rigorously, $z=1-0$ ) corresponds to the horizon ($R = 2 \mu$).

Equations (\ref{i2.1}) with  conditions of the finiteness on the horizon (\ref{i2.1c}) imposed
 imply the following  integral of motion:
  \bear{i2.1d}
 \frac{1}{2}(1 -z)  \sum_{s,l =1,2} h_s A_{sl} \frac{d y^s}{dz} \frac{d y^l}{dz}
    +  \sum_{s =1,2} h_s \frac{d y^s}{dz}
              \\ \nonumber
           -  \sum_{s = 1,2} q_s^2 \exp(- \sum_{l =1,2} A_{sl} y^l)   = 0.
 \ear
Equations (\ref{i2.1}) and (\ref{i2.1c}) appear for  special
solutions to Toda-type equations  \cite{IMp2,IMp3,IMtop}
\beq{i2.1T}
  \frac{d^2 z^s}{du^2}   =    K_s Q_s^2 \exp(\sum_{l =1,2} A_{sl} z^l ),
 \eeq
for  functions
\beq{i2.z}
z^s(u) = - y^s -   \mu b^s u,
\eeq
$s = 1,2$, depending
on the harmonic radial variable $u$: $\exp(- 2 \mu u) = 1 -z$,
with  the following asymptotical
behavior for $u \to + \infty$ (on the horizon) imposed:
  \beq{i2.1as}
  z^s(u) = - \mu b^s u + z^s_{0} + o(1),
  \eeq
  where $z_{s0}$ are constants,  $s = 1,2$. Here and in the following we denote
  \beq{i2.b}
   b^s = 2 \sum_{l =1,2} A^{sl},
  \eeq
  where the inverse matrix $(A^{sl}) =  (A_{sl})^{-1}$ is well defined due
  to $\lambda_1 \neq - \lambda_2$.
  This follows from the relations
   \beq{i2.B}
      A_{sl} = 2 B_{sl} h_l, \qquad  B_{sl} = \frac{1}{2}
       + \varepsilon \chi_s \chi_l \lambda_s \lambda_l,
     \eeq
   where $\chi_1 = +1 $, $\chi_2 = - 1 $
   and the invertibility of the matrix $(B_{sl})$ for $\lambda_1 \neq - \lambda_2$,
   due to the relation $\det (B_{sl}) = \frac{1}{2} \varepsilon (\lambda_1 + \lambda_2)^2$.

   The  energy  integral of motion for (\ref{i2.1T}), which is compatible with the asymptotic
 conditions (\ref{i2.1as}),
 \bear{i2.1ET}
   E = \frac{1}{4} \sum_{s,l =1,2} h_s A_{sl} \frac{d z^s}{du} \frac{d z^l}{du}
            \\ \nonumber
     - \frac{1}{2} \sum_{s=1,2}  Q_s^2 \exp(\sum_{l =1,2} A_{sl} z^l )  =
     \frac{1}{2} \mu^2 \sum_{s=1,2} h_s b^s,
  \ear
   leads  to eq.  (\ref{i2.1d}).

{\bf Remark 2.} {\em The derivation of the solutions (\ref{i.3})-(\ref{i.3bm}), (\ref{i3.1})-(\ref{i3.1b})
 may be extracted from the  relations of \cite{IMp1,IMp2,IMp3}, where the solutions with a horizon were obtained from 
 general spherically symmetric solutions governed by Toda-like  equations. These  Toda-like  equations
 contain a non-trivial part corresponding to a non-degenerate  (quasi-Cartan) matrix $A$. In our case 
 these equations are given by (\ref{i2.1T}) with the matrix $A$ from (\ref{i2.B}) and
 the condition   ${\rm det} A \neq 0$   implies $\lambda_1 \neq - \lambda_2$. The master equations
 (\ref{i3.1}) are equivalent to these Toda-like  equations.
 Fortunately,  for  $\lambda_1 = - \lambda_2$ and $\varepsilon = +1$ the solution does exist. It obeys  eqs. 
 (\ref{i.3})-(\ref{i.3bm}) and (\ref{i3.1})-(\ref{i3.1b}) with  $H_i = H^{\frac{Q_i^2}{Q_1^2 + Q_2^2}}$,
 $i = 1,2$, where $H = 1 + \frac{P}{R}$ and $P > 0$ satisfies  $P(P + 2\mu) = K_1 (Q_1^2 + Q_2^2)$,
 $K_1 > 0$. 
  For $\lambda_1 = - \lambda_2$ the solution reads:
 $$ds^2 = H^{h_1} \biggl\{ - H^{-2 h_1} \left( 1 - \frac{2\mu}{R} \right)
 dt^2  +  \frac{dR^2}{1 - \frac{2\mu}{R}} + R^2  d \Omega^2_{2}   \biggr\},$$
 $$\exp(\varphi)=   H^{h_1 \lambda_1 \varepsilon}, \quad
 F^{(1)}=  \frac{Q_1}{R^2}   H^{-2}  dt \wedge dR, \quad F^{(2)}  =  Q_2 \tau. $$
 We have verified this solution by using MATHEMATICA. It is also valid for $\varepsilon = -1$ and 
 $\lambda_1^2 < \frac{1}{2}$. } 

\section{Some integrable cases}

Explicit analytical solutions
to  eqs. (\ref{i3.1}), (\ref{i3.1a}), (\ref{i3.1b}) do not exist.
One may try to seek the solutions in the form

\beq{i3.12}
  H_{s} = 1 + \sum_{k = 1}^{\infty} P_s^{(k)}
  \left(\frac{1}{R}\right)^k,
\eeq
where $P_s^{(k)}$ are constants, $k = 1,2,\ldots, $ and
$s =1,2$, but only in few integrable cases the chain of
 equations for $P_s^{(k)}$ is dropped.

For $\varepsilon = + 1$,  there exist at least four integrable configurations
related to the Lie algebras $A_1 + A_1$, $A_2$,  $B_2 = C_2$ and $G_2$.

\subsection{$(A_1 + A_1)$-case}

Let us consider the case $\varepsilon = 1$ and
\beq{i4.0}
    \left(A_{ss'}\right)=
  \left( \begin{array}{*{6}{c}}
     2 &  0\\
     0 & 2\\
\end{array}
\right) .
\eeq
We obtain
\beq{i4.1}
  \lambda_1 \lambda_2 = \frac{1}{2}.
\eeq

For $\lambda_1 = \lambda_2$
we get a dilatonic coupling corresponding to string induced
model. The matrix (\ref{i4.0}) is the
Cartan matrix for the Lie algebra $A_1 + A_1$ ($A_1 = sl(2)$).
In this case
\beq{i4.2}
 H_s = 1 + \frac{P_s}{R},
\eeq
where
\beq{i4.3}
 P_s (P_s + 2 \mu) = K_s Q_s^2,
\eeq
$s = 1,2$. For   positive roots
of (\ref{i4.3})
\beq{i4.3p}
    P_s  = P_{s+} =  - \mu + \sqrt{\mu^2 + K_s Q^2_s},
\eeq
we are led to a well-defined solution for $R > 2\mu$  with asymptotically
flat metric  and  horizon at $R = 2 \mu$. We note that in the case
$\lambda_1 = \lambda_2$ the  $(A_1 + A_1)$-dyon
solution has a composite analog which was considered earlier in \cite{GM,ChHsuL};
see also \cite{Br0} for  certain generalizations.

\subsection{$A_2$-case}

Now we put $\varepsilon = 1$ and
\beq{i4.4a}
    \left(A_{ss'}\right)=
  \left( \begin{array}{*{6}{c}}
     2 &  -1\\
     -1 & 2\\
\end{array}
\right).
\eeq
We get
\beq{i4.4}
 \lambda_1 = \lambda_2 = \lambda, \qquad  \lambda^2 = 3/2. \qquad
\eeq
This value of dilatonic coupling constant appears after reduction to four dimensions of the 5d
Kaluza-Klein model. We get $h_s = 1/2$ and (\ref{i4.4a})
is the Cartan matrix for the Lie algebra $A_2 = sl(3)$.
In this case we obtain \cite{IMp2}

\beq{i4.5}
H_s = 1 + \frac{P_s}{R} + \frac{P_s^{(2)}}{R^2},
\eeq
where
\bear{i4.6}
 2 Q_s^2 = \frac{P_s (P_s + 2 \mu) (P_s + 4 \mu)}{P_1 + P_2 + 4 \mu},
 \\ \label{i4.6a}
 P_s^{(2)} = \frac{P_s (P_s + 2 \mu) P_{\bar{s}}}{2 (P_1 + P_2 + 4 \mu)},
\ear
$s = 1,2$ ($\bar{s} = 2,1$).

In the composite case \cite{ABDI} the Kaluza-Klein uplift to $D=5$ gives us
the well-known Gibbons-Wiltshire solution  \cite{GibW},
which follows from the general spherically symmetric
dyon solution (related to $A_2$ Toda chain) from ref.  \cite{Lee}.

\subsection{$C_2$ and $G_2$ cases}

If we put $\varepsilon = 1$ and
\beq{i4.7a}
    \left(A_{ss'}\right)=
  \left( \begin{array}{*{6}{c}}
     2 &  -1\\
     -k & 2\\
    \end{array} \right)
      \quad  {\rm or}   \quad    
       \left(A_{ss'}\right)=
       \left( \begin{array}{*{6}{c}}
          2 &  -k\\
          -1 & 2\\
\end{array}
\right),
\eeq
we also get integrable configurations for $k = 2, 3$, corresponding
to the Lie algebras $B_2 = C_2$ and $G_2$, respectively, with the degrees of polynomials $(3,4)$
and $(6,10)$. From (\ref{i.16a}),   (\ref{i.16b}) and (\ref{i4.7a}) 
 we get the following relations for the dilatonic couplings:
\beq{i4.7aa}
 \frac{1}{2} +  \lambda_2^2 = k \left(\frac{1}{2} +  \lambda_1^2 \right),
 \quad  1 - 2 \lambda_1 \lambda_2  = - \frac{1}{2} -  \lambda_2^2,
 \eeq
 or 
\beq{i4.7ab}
 \frac{1}{2} +  \lambda_1^2 = k \left(\frac{1}{2} +  \lambda_2^2 \right),
 \quad  1 - 2 \lambda_1 \lambda_2  = - \frac{1}{2} -  \lambda_1^2.
 \eeq

Solving eqs. (\ref{i4.7aa}) we get 
$(\lambda_1, \lambda_2) = \pm (\sqrt{2}, \frac{3}{\sqrt{2}})$ for $k =2$ and 
$(\lambda_1, \lambda_2) = \pm \left(\frac{5}{\sqrt{6}}, 3\sqrt{\frac{3}{2}}\right)$ for $k =3$.
The solution to eqs. (\ref{i4.7ab}) is given  by permutation of  $\lambda_1$ and $\lambda_2$.

The exact black hole (dyonic-like) solutions for Lie algebras $B_2 = C_2$ and $G_2$ 
will be analyzed in detail in  separate publications. 
They do not exist for the case $\lambda_1 = \lambda_2$. We note that for the $B_2 = C_2$ case ($k =2$)
the polynomials $H_i$, $i =1,2$, were calculated in \cite{GrIK}.

\subsection{Special solution with two dependent charges}

There exists also a special solution
\beq{i4.7}
 H_s = \left(1 + \frac{P}{R}\right)^{b^s},
\eeq
with $P > 0$ obeying
\beq{i4.8a}
  \frac{K_s}{b_s} Q_s^2 = P (P + 2 \mu),
\eeq
 $s = 1,2$.
 Here  $b^s \neq 0$ is defined in (\ref{i2.b}).
 This solution is a special case of more general
 ``block orthogonal'' black brane  solutions  \cite{Br,IMJ2,CIM}.

 The calculations give us the following relations:
 \beq{i4.8b}
  b^s = \frac{2 \lambda_{\bar s}}{\lambda_1 + \lambda_2} K_s,
 \eeq

 \beq{i4.8c}
    Q_s^2 \frac{(\lambda_1 + \lambda_2)}{2 \lambda_{\bar s}}
    = P (P + 2 \mu) = \frac{1}{2} Q^2,
 \eeq
 where $s = 1, 2$ and ${\bar s} = 2,1$, respectively.
 Our solution is well defined if $\lambda_1 \lambda_2 > 0$,
 i.e. the two  coupling constants have the same sign.

 For positive roots of (\ref{i4.8c})
\beq{i4.8p}
    P  = P_{+} =  - \mu + \sqrt{\mu^2 + \frac{1}{2}Q^2}
\eeq
we get for $R > 2\mu$ a well-defined  solution with asymptotically
flat metric  and   horizon at $R = 2\mu$.
It should be noted that this special solution is valid for
both signs $\varepsilon = \pm 1$.
We have
\bear{i4.9}
 &&ds^2 =  H^{2}\biggl\{ -  H^{-4} \left( 1 - \frac{2\mu}{R} \right) dt^2
  +  \frac{dR^2}{1 - \frac{2\mu}{R}} + R^2  d \Omega^2_{2} \biggr\},
 \\  \label{i4.10}
 &&\varphi = 0, \qquad \qquad \qquad \qquad \qquad \qquad \qquad
 \\  \label{i4.10em}
 &&F^{(1)}= \frac{Q_1}{H^2 R^2} dt \wedge dR, \qquad
  F^{(2)}  =  Q_2 \tau,
\ear
where $H = 1 + \frac{P}{R}$ with $P$ from  (\ref{i4.8p})
and
\beq{i4.11}
  Q_1^2 = \frac{\lambda_{2}}{\lambda_1 + \lambda_2} Q^2, \qquad
  Q_2^2 = \frac{ \lambda_{1}}{\lambda_1 + \lambda_2} Q^2.
\eeq

By changing the radial variable, $r = R + P$, we get
 \bear{i4.12}
  &&ds^2 =   - f(r) dt^2 +   f(r)^{-1} dr^2 + r^2  d \Omega^2_{2},
     \\    \label{i4.12F}
  &&F^{(1)}= \frac{Q_1}{r^2} dt \wedge dr, \quad F^{(2)}  =  Q_2 \tau,
  \quad \varphi = 0,
    \ear
 where $f(r) = 1 - \frac{2GM}{r} + \frac{Q^2}{2r^2}$, $ Q^2 = Q_1^2 + Q_2^2$ and $GM = P + \mu$ =
 $\sqrt{\mu^2 + \frac{1}{2} Q^2} > \frac{1}{\sqrt{2}} |Q|$.

 The metric in these variables is coinciding with the well-known Reissner-Nordstr\"om  
 metric governed by two  parameters: $GM > 0$ and $ Q^2 < 2 (GM)^2$. 
 We have two horizons in this case.  Electric and magnetic charges are not independent but obey eqs.
 (\ref{i4.11}).

\subsection{The limiting $A_1$-cases}

In the following we will use two limiting solutions: an electric one
with $Q_1 = Q \neq 0$ and   $Q_2 = 0$,
\bear{i4.e}
H_1 = 1 + \frac{P_1}{R}, \qquad H_2 = 1,
\ear
and a magnetic one  with $Q_1 = 0$ and $Q_2 = Q \neq 0$,
 \bear{i4.m}
H_1 = 1, \qquad  H_2 = 1 + \frac{P_2}{R}.
 \ear
In both cases $P_s  =   - \mu + \sqrt{\mu^2 + K_s Q^2}$.
These solutions correspond to  the Lie algebra $A_1$.
In various notations the solution  (\ref{i4.e}) appeared earlier in \cite{BronShikin,Hein,GM}, 
and it was extended to the multidimensional
case in  \cite{Hein,GM,BBFM,BI}. The special case with $\lambda^2_1 = 1/2$, $\varepsilon= 1$,
was considered earlier in \cite{Gibbons,GHS}.

\section{Physical parameters}

Here we consider certain physical parameters
corresponding to the solutions under consideration.

\subsection{Gravitational mass and scalar charge}

For ADM gravitational mass we get from (\ref{i.3})
\beq{i5.1}
 GM =   \mu +  \frac{1}{2} (h_1 P_1 + h_2 P_2),
\eeq
where  the parameters $P_s = P_s^{(1)}$ appear in
eq. (\ref{i3.12}) and $G$ is the gravitational constant.

The scalar charge just follows  from (\ref{i.3a}):
\beq{i5.1s}
 Q_{\varphi} =  \varepsilon (\lambda_1 h_1 P_1 -  \lambda_2 h_2 P_2).
\eeq

For the special solution (\ref{i4.7})  with $P >0$ we get

\beq{i5.1sim}
  GM =  \mu + P = \sqrt{\mu^2 + Q^2}, \qquad
  Q_{\varphi} = 0.
 \eeq

For fixed charges $Q_s$ and the extremality parameter $\mu$
the mass $M$ and scalar charge $Q_{\varphi}$ are not
independent but obey a certain constraint. Indeed,
for fixed parameters $P_s = P_s^{(1)}$ in   (\ref{i3.12}) we
get
\beq{i5.12}
      y^s = \ln H_s = \frac{P_s}{2\mu} z + O(z^2),
 \eeq
for   $z \to + 0$, which after substitution into  (\ref{i2.1d}) gives  (for $z =0$)
the following identity:

\beq{i5.1p}
 \frac{1}{2} \sum_{s,l =1,2} h_s A_{sl} P_s P_l
 + 2 \mu \sum_{s =1,2} h_s P_s  =  \sum_{s =1,2} Q_s^2.
\eeq

By using  eqs. (\ref{i5.1}) and (\ref{i5.1s}) this identity may be rewritten
in the following form:

\beq{i5.1id}
     2 (GM)^2   +   \varepsilon  Q_{\varphi}^2   = Q_1^2 + Q_2^2 + 2 \mu^2.
\eeq

It is remarkable that this formula does not contain $\lambda$.
We note that in the extremal case $\mu = +0$ this relation for $\varepsilon = 1$
 was obtained earlier  in \cite{PTW}.

\subsection{The Hawking temperature and  entropy}

The Hawking temperature corresponding to
the solution is found to be
  \beq{i5.2}
 T_H=   \frac{1}{8 \pi \mu}  H_{10}^{- h_1} H_{20}^{- h_2},
 \eeq
where $H_{s0}$ are defined in (\ref{i3.1a}).
Here and in the following we put $c= \hbar = \kappa =1$.

For special solutions (\ref{i4.7})   with $P >0$ we get
\beq{i5.3sim}
  T_H =  \frac{1}{8 \pi \mu} \left(1 + \frac{P}{2 \mu}\right)^{-2}.
 \eeq
In this case the Hawking temperature $T_H$ does not depend upon
 $\lambda_s$ and  $\varepsilon$, when $\mu$ and $P$ (or $Q^2$) are fixed.

The Bekenstein-Hawking (area) entropy $S = A/(4G)$,
corresponding to the horizon at $R = 2\mu$, where $A$ is the horizon area,  reads
\beq{i5.2s}
S_{BH} =   \frac{4 \pi \mu^2}{G}  H_{10}^{h_1} H_{20}^{h_2}.
\eeq
It follows from (\ref{i5.2}) and (\ref{i5.2s}) that the product
\beq{i5.2st}
  T_H  S_{BH} =   \frac{\mu}{2G}
\eeq
does not depend upon  $\lambda_s$,  $\varepsilon$ and the charges $Q_s$.
This product does not use an explicit form of the moduli functions $H_s(R)$.

\subsection{PPN parameters}

Introducing a new radial variable $\rho$ by the relation
 $R =   \rho (1 + (\mu/2\rho))^2$
($\rho > \mu/2$),  we obtain  the 3-dimensionally
conformally flat form of the metric  (\ref{i.3})
\bear{i5.3}
g =   U \Biggl\{ -
U_1 \frac{\left(1 - (\mu/2\rho) \right)^2}
{\left(1 + (\mu/2\rho) \right)^2} dt \otimes dt +
\left(1 + \frac{\mu}{2 \rho} \right)^4
\delta_{ij} dx^i \otimes dx^j \Biggr\},
\ear
where  $\rho^2 =  |x|^2 =   \delta_{ij}x^i x^j$ ($i,j =   1,2,3$)
and
\beq{5.5.1a}
U =   \prod_{s = 1,2} H_s^{h_s},\qquad U_1 =   \prod_{s  = 1,2} H_s^{-2 h_s}.
\eeq

The parametrized post-Newtonian (PPN) parameters
$\beta$ and $\gamma$ are defined by the following standard relations:
\bear{A.1}
g_{00} =   - (1 -  2 V + 2 \beta V^2 ) + O(V^3),
\\
\label{A.2}
g_{ij} =   \delta_{ij}(1 + 2 \gamma V ) + O(V^2),
\ear
$i,j =   1,2,3$, where $V =   GM/\rho$
is  Newton's potential, $G$ is the gravitational constant and
$M$ is the gravitational mass (for our case see (\ref{i5.1})).

The calculations of PPN (or Eddington) parameters for the metric (\ref{i5.3})
give the same result as in \cite{FIMS}:
\beq{i5.8}
\beta  = 1 +   \frac{1}{4(GM)^2}  (Q_1^{2} + Q_2^{2}), \qquad \gamma = 1.
\eeq

These parameters  do not depend upon $\lambda_s$
and $\varepsilon$. They may be calculated just without knowledge of the explicit
relations for  the moduli functions $H_s(R)$.

 These parameters (at least  formally)
obey  the observational restrictions for the solar system \cite{Wil},
when  $Q_s/(2GM)$ are small enough.

\section{Bounds on  mass and scalar charge}

Here we outline the following hypothesis, which is supported by certain numerical calculations \cite{ABDI,ABIM}.
For $h_1 = h_2$ this conjecture was proposed in ref. \cite{ABDI}.

{\bf Conjecture.}
{\em For any $h_1 >0$, $h_2 >0$, $\varepsilon = \pm 1$, $Q_1 \neq 0$, $Q_2\neq 0$ and $\mu > 0$:
(A) the moduli functions $H_s(R)$, which obey  (\ref{i3.1}), (\ref{i3.1a}) and (\ref{i3.1b}),
are uniquely defined and hence the parameters $P_1$, $P_2$, the gravitational mass $M$ and
the scalar charge $Q_{\varphi}$ are uniquely defined too;
(B) the parameters $P_1$, $P_2$ are positive and the  functions $P_1 = P_1(Q_1^2,Q_2^2)$,
 $P_2 = P_2(Q_1^2,Q_2^2)$  define a diffeomorphism of $\R_{+}^2$ ($\R_{+} = \{x| x>0 \}$);
  (C) in the limiting case we have: (i) for $Q_2^2 \to + 0$:
  $P_1 \to   - \mu + \sqrt{\mu^2 + K_1 Q_1^2}$, $P_2 \to +0$
  and (ii) for $Q_1^2 \to + 0$: $P_1 \to +0$, $P_2 \to   - \mu + \sqrt{\mu^2 + K_2 Q_2^2}$.}

 The  conjecture could be readily verified  for the case  $\varepsilon = 1$,
$\lambda_1 \lambda_2 = 1/2$.
 Another integrable case  $\varepsilon = 1$, $\lambda_1 = \lambda_2 = \lambda$, 
 $\lambda^2 = 3/2$ is more involved  \cite{ABIM}.

 The conjecture implies the following proposition.

{\bf Proposition 1.}
{\em For $h_s > 0$, $Q_s \neq 0$, $\lambda_s \neq 0$ ($s=1,2$) and $\lambda_1 + \lambda_2 \neq 0$ 
we have the following   bounds on the gravitational mass $M$  and the scalar charge $Q_{\varphi}$:

\bear{i5.13p}
   \mu  + \frac{h_{min}}{2}\left(- \mu + \sqrt{h^{-1}_{min} (Q_1^2 + Q_2^2) + \mu^2} \right) < GM
    \leq \sqrt{\frac{1}{2} (Q_1^2 + Q_2^2) + \mu^2}, \\     \label{i5.4p}
   |Q_{\varphi}| < |\lambda|_{max } h_{min}
   \left(- \mu + \sqrt{h^{-1}_{min} (Q_1^2 + Q_2^2)+ \mu^2}\right),
    \qquad     \quad
  \ear
for $\varepsilon = +1$ $(0< h_s < 2)$ and
\bear{i5.13m}
 \sqrt{\frac{1}{2} (Q_1^2 + Q_2^2) + \mu^2} \leq GM
       < \mu  + \frac{h_{max}}{2} \left(- \mu + \sqrt{h_{max}^{-1} (Q_1^2 + Q_2^2) + \mu^2}\right),
      \\     \label{i5.4m}
  |Q_{\varphi}| < |\lambda|_{max } h_{max}
     \left(- \mu + \sqrt{h^{-1}_{max} (Q_1^2 + Q_2^2)+ \mu^2} \right),
      \qquad       \quad
     \ear
for $\varepsilon = -1$  $(h_s > 2)$. Here
$h_{min} = {\rm min} (h_1,h_2)$, $h_{max} = {\rm max} (h_1,h_2)$, and
  $|\lambda|_{max} = { \rm max } (|\lambda|_{1}, |\lambda|_{2} )$;
  $h_{min} = (\frac{1}{2} +   |\lambda|^2_{max})^{-1}$ for $\varepsilon = +1$ and
  $h_{max} = (\frac{1}{2} -   |\lambda|^2_{max})^{-1}$ for $\varepsilon = -1$ }.

  Here we illustrate the bounds on $M$ and $Q_{\varphi}$ graphically by four figures,
which represent a  set of physical parameters  $GM$ and $Q_{\varphi}$ for
 $Q_1^2 + Q_2^2 = Q^2 =2$ and $\mu = 1$.

The left panel of Fig. 1  corresponds to the case $\varepsilon = +1$,
 $\lambda_1= \sqrt{ \frac{1}{2}}$ and $\lambda_2 = 1/2$,
 while the right panel of this figure describes  the case   $\varepsilon = +1$,
 $\lambda_1= \sqrt{ \frac{1}{2}}$ and $\lambda_2 = -1/4$.

On Fig. 2 the left panel illustrates the  case $\varepsilon = -1$,
$\lambda_1 = \sqrt{0.499}$ and $\lambda_2 = 1/2$,
while the right panel represents  the  case $\varepsilon = -1$, $\lambda_1 = \sqrt{0.499}$ and
  $\lambda_2 = -1/4$.

Two arcs on the left panels of Figs. 1 and 2 contain the  points  with $Q_{\varphi} =0$ corresponding to the special solution from Sect. 3.4.

\begin{figure}
\begin{center}
\includegraphics[width=0.46\textwidth]{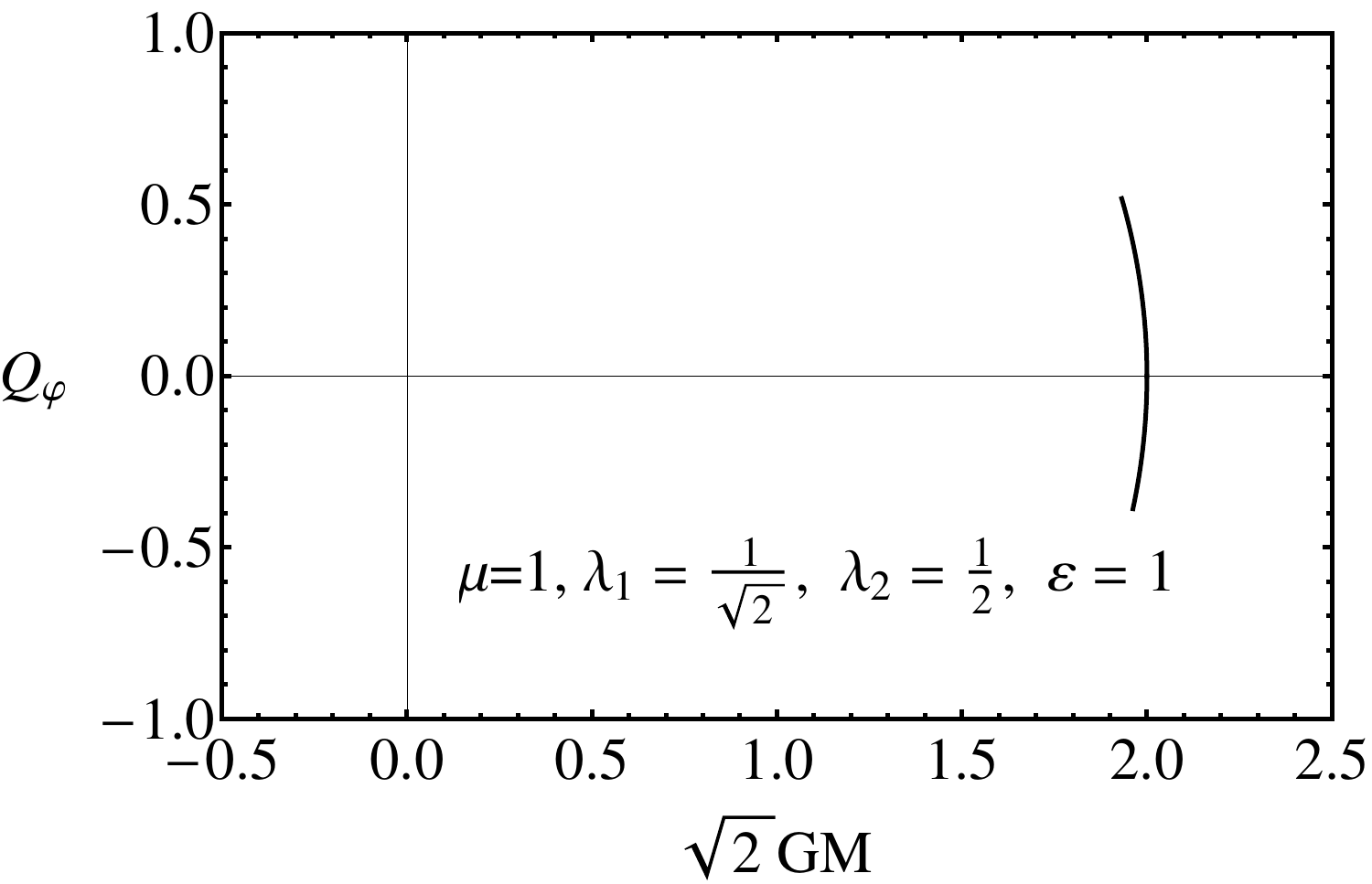}\qquad
\includegraphics[width=0.46\textwidth]{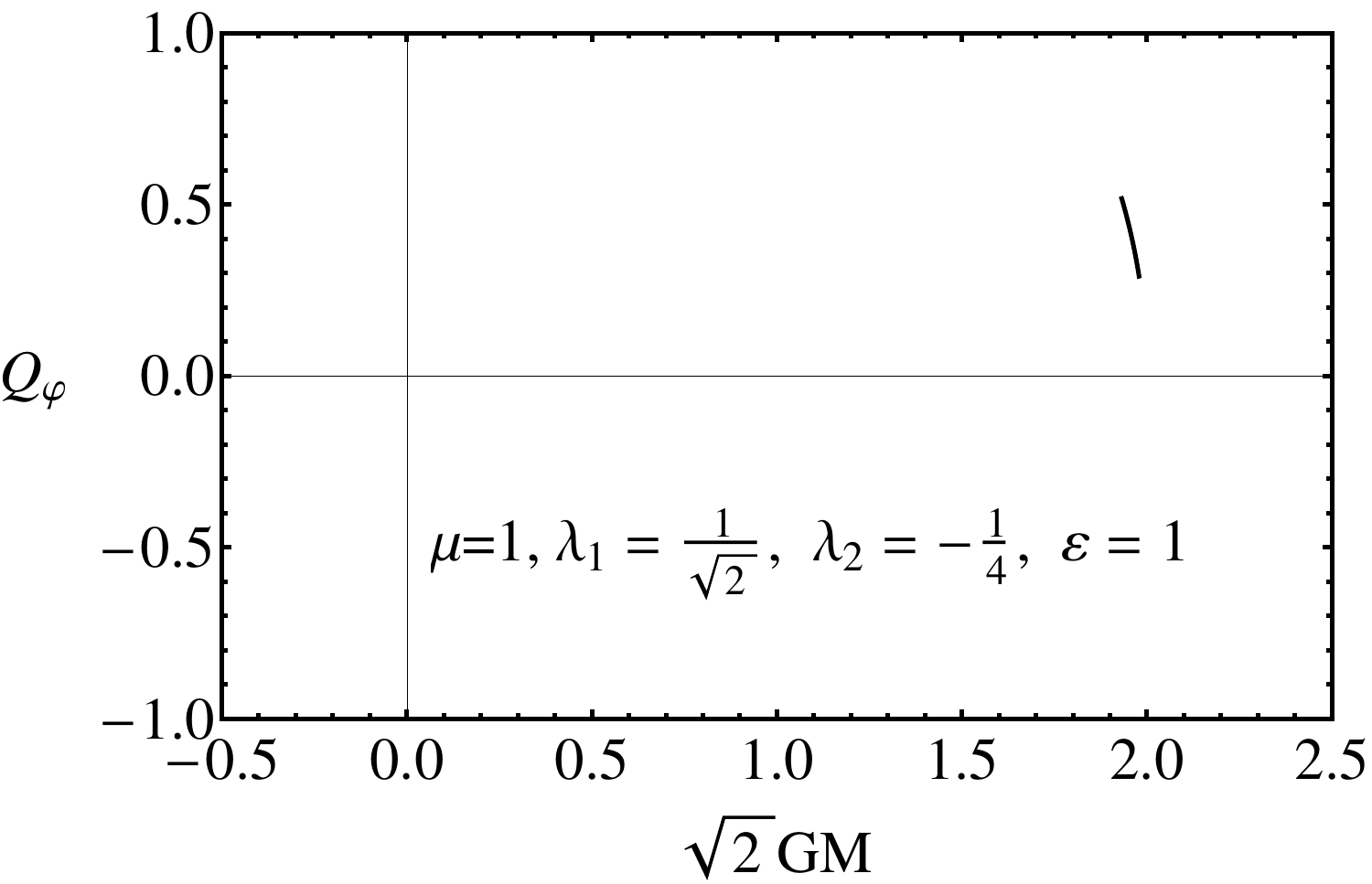}
\caption{Graphical illustration of  bounds on $M$ and $Q_{\varphi}$ for  $\varepsilon = 1$, $\lambda_1= 1/\sqrt{2}$,  $\mu = 1$ and  $Q_1^2 + Q_2^2 = 2$. The only difference between two diagrams is $\lambda_2= 1/2$ (left panel) and $\lambda_2= - 1/4$ (right panel).}\label{f1}
\end{center}
\end{figure}

\begin{figure}
\begin{center}
\includegraphics[width=0.46\textwidth]{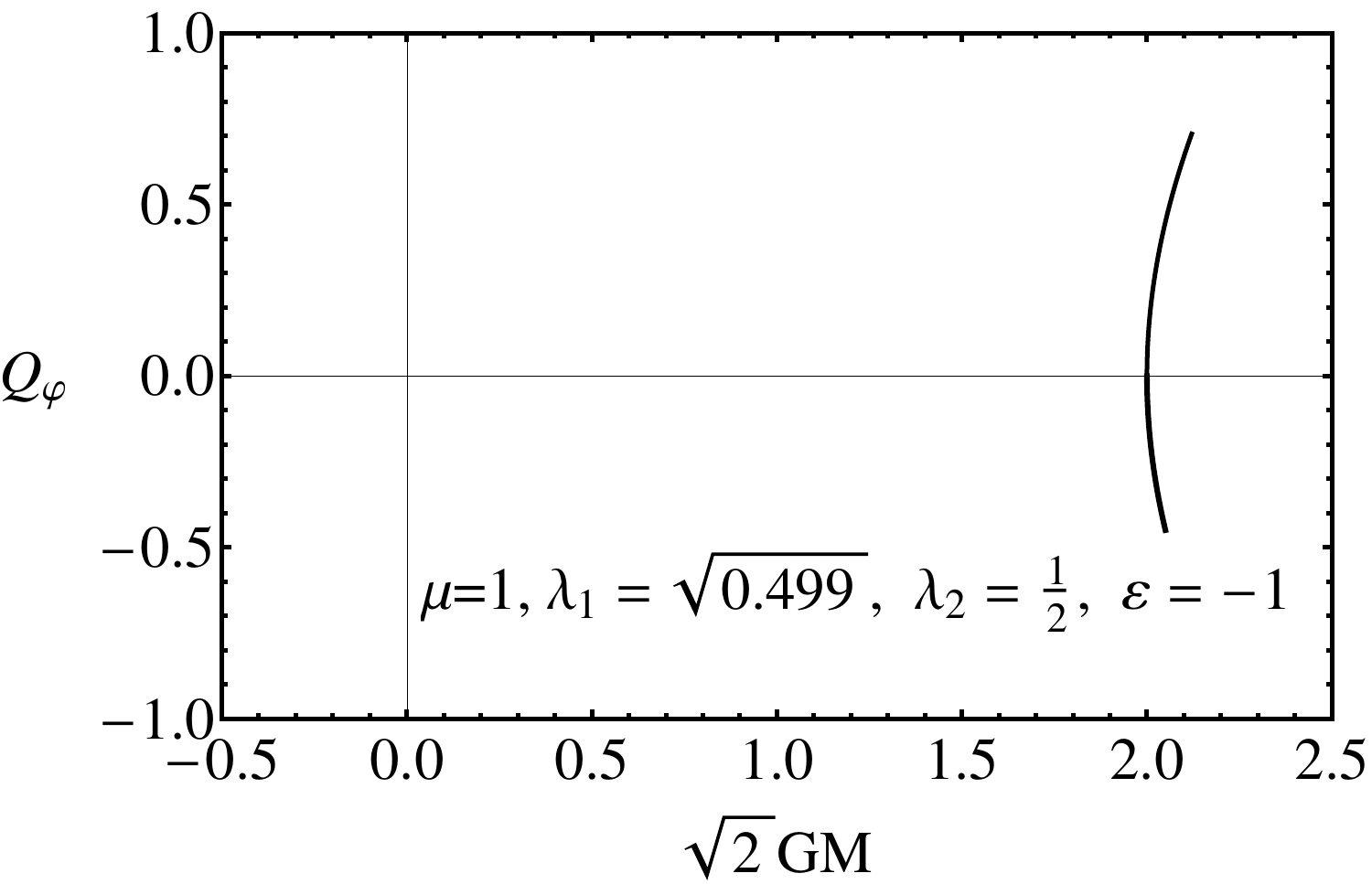}\qquad
\includegraphics[width=0.46\textwidth]{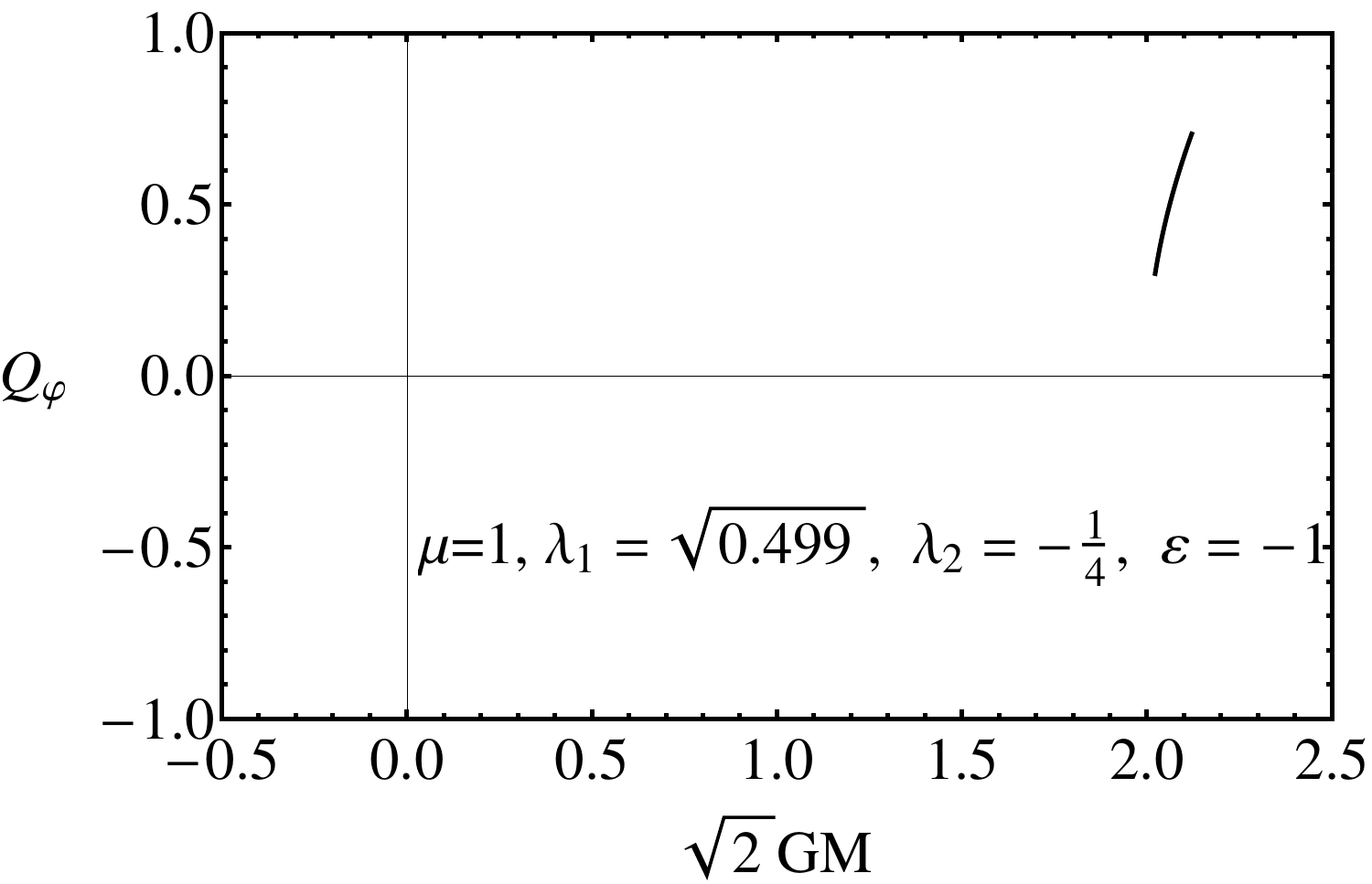}
\caption{Graphical illustration of  bounds on $M$ and $Q_{\varphi}$
for  $\varepsilon = -1$, $\lambda_1= \sqrt{0.499 }$, $\mu = 1$ and  $Q_1^2 + Q_2^2 = 2$.
The only difference between two diagrams is $\lambda_2= 1/2$ (left panel) and $\lambda_2= - 1/4$ (right panel).}\label{f2}
\end{center}
\end{figure}

In proving Proposition 1 we use the following  lemma.

{\bf Lemma.} {\em Let
\beq{i5.17}
f(\mu,h;Q^2) = \mu  + \frac{h}{2} \left(- \mu + \sqrt{h^{-1} Q^2 + \mu^2}\right),
\eeq
be a function of two variables  $\mu >0$ and $h > 0$ with fixed value of $Q^2 > 0$. 
Then:
(i) for fixed value of $\mu$ the function
 $f(\mu,h;Q^2)$ is monotonically increasing with respect to $h$;
(ii) for fixed value of $h \in (0,2)$ the function $f(\mu,h;Q^2)$ is 
monotonically increasing with respect to $\mu$ and 
$f(+0,h;Q^2) = \frac{1}{2} \sqrt{h Q^2} < f(\mu,h;Q^2)$. }

The proof of the lemma is trivial: item (i) just follows from the identity
\beq{i5.18}
f(\mu,h;Q^2) = \mu  +  \frac{Q^2}{2 (\mu + \sqrt{h^{-1} Q^2 + \mu^2})},
\eeq
while item (ii) could be readily verified by using  the relation
\beq{i5.19}
\frac{\partial f}{\partial \mu} = 1  + \frac{h}{2}
 \left( -1 +  \frac{\mu}{ \sqrt{h^{-1} Q^2 + \mu^2}} \right) > 0
\eeq
for $h \in (0,2)$.

{\bf Proof of Proposition 1.}
Let us prove the relations (\ref{i5.13p}), (\ref{i5.4p}), (\ref{i5.13m})
and (\ref{i5.4m}) using the  conjecture.
The right inequality (or equality) in (\ref{i5.13p}) just follows
from the eq.  (\ref{i5.1id}), while the left inequality (or equality) in (\ref{i5.13m}) follows
from   (\ref{i5.1id}) and $M > 0$ which is valid due to eq. (\ref{i5.1}), $h>0$ and the inequalities
$P_1 >0$, $P_2 > 0$ (due to the conjecture.).

Now let us verify the left inequality in (\ref{i5.13p}). We fix the charges by the relation
$Q_1^2 + Q_2^2 = Q^2$, $Q >0$, and put $Q_1^2 = \frac{1}{2} Q^2(1+ x)$, $Q_2^2 = \frac{1}{2} Q^2(1 - x)$, where
$-1 < x < 1$. Due to (\ref{i5.1id}) and $M > 0$ we can use the following parametrization:
   \beq{i5.14}
         \sqrt{2}GM = R\cos{\psi}, \quad  Q_{\varphi} = R \sin{\psi}, \qquad   R = \sqrt{Q^2 + 2 \mu^2},
     \eeq
where $|\psi| < \pi/2$. Due  to the conjecture and relations (\ref{i5.1}), (\ref{i5.1s})
 we see that $\psi = \psi(x)$ is a smooth function which obeys
    \beq{i5.15}
         \psi(1 - 0) =  \psi_1, \qquad \psi( - 1 + 0) =  \psi_2 .
      \eeq
Here  $R \cos{\psi_i} = \sqrt{2} ( \mu +  \frac{h_i}{2} P_i )$ and
$R \sin{\psi_i} = \lambda_i h_i P_i \chi_i$, where
 $P_i = -\mu +  \sqrt{K_i Q^2+ \mu^2}$, ($K_i = h_i^{-1}$) $i =1,2$, and $\chi_1 = 1$, $\chi_2 =  -1$.

 We put $\lambda_1 > 0$ without loss of generality.
 The limit $x \to +1 -0$ corresponds to a pure electric black hole while the limit
 $x \to - 1 +0$ corresponds to a pure magnetic one.

 To prove eqs. (\ref{i5.13p})   and (\ref{i5.4p})
 one should verify the inequality
     \beq{i5.15a}
         \psi_2 < \psi(x) <  \psi_1
      \eeq
for all $x \in (-1,1)$. Indeed, due to relations  (\ref{i5.15a}) the  points $(\sqrt{2}GM,Q_{\varphi})$
describe an open arc in the circle (see Fig. 1). One of the endpoints of this arc with $\psi =  \psi_{i_0}$,
$i_0 =1,2$, gives us the lower bound for $GM$ and upper bound for $|Q_{\varphi}|$. Due to the lemma this point
 corresponds to $i_0$ obeying $h_{i_0} =  h_{min} = {\rm min} (h_1,h_2)$,  $P_{i_0} = -\mu +  \sqrt{K_{i_0} Q^2+ \mu^2}$  and $P_{\bar{{i_0}}} = 0$.

Let us suppose that (\ref{i5.15a}) is not valid. Without loss of generality we
 put   $\psi(x_{*}) \geq \psi_1$   for some $x_{*}$. Then,  using  (\ref{i5.15}) and the smoothness of
the function $\psi(x)$, we get that for some $x_1 \neq x_2$:  $\psi(x_{1}) = \psi(x_{2})$.
This follows from the intermediate value theorem which  states that
if  $f(x)$ is a continuous function on the interval $[a, b]$, then, for any  $d \in [f (a), f (b)]$, 
 there is a point
$c \in   [a, b]$ such that $f(c) = d$. (Here for $f(a) > f(b)$, $[f (a), f (b)]$ is meant to mean $[f (b), f (a)]$.)
Hence for two different
sets  $(Q_1^2,Q_2^2)_1 \neq (Q_1^2,Q_2^2)_2$ we obtain the same coinciding sets: $(GM,Q_{\varphi})_1 = (GM,Q_{\varphi})_2$
and hence  $(P_1,P_2)_1 = (P_1,P_2)_2$; see (\ref{i5.1}),  (\ref{i5.1s}) and $\lambda_1 \neq - \lambda_2$. But due to our  conjecture the map   $(Q_1^2,Q_2^2) \mapsto (P_1,P_2)$ is bijective (i.e. it is one-to-one correspondence). This implies   $(P_1,P_2)_1 \neq (P_1,P_2)_2$.
We get a contradiction which proves our proposition for $\varepsilon = 1$ and arbitrary $Q_1^2 + Q_2^2 > 0$.

The proofs of the right inequality in   (\ref{i5.13m})  and the bound  (\ref{i5.4m}) for $\varepsilon = - 1$ are quite similar to that for $\varepsilon =  1$.
 The only difference here is the use of parametrization
 \beq{i5.14m}
         \sqrt{2}GM = R\cosh{\psi}, \quad  Q_{\varphi} = R \sinh{\psi},
         \qquad   R = \sqrt{Q^2 + 2 \mu^2},
     \eeq
instead of  (\ref{i5.14}). Due to relations  (\ref{i5.15a}) the  points $(\sqrt{2}GM,Q_{\varphi})$
describe an open arc in the hyperbola (see Fig. 2). One of the endpoints of this arc with $\psi =  \psi_{j_0}$,
$j_0 =1,2$, gives us the upper bound for $GM$ and the upper bound for $|Q_{\varphi}|$. Due to the lemma this point
 corresponds to $j_0$ obeying $h_{j_0} =  h_{max} = {\rm max} (h_1,h_2)$,  $P_{j_0} = -\mu +  \sqrt{K_{j_0} Q^2+ \mu^2}$  and $P_{\bar{{j_0}}} = 0$.
Thus, Proposition 1 is proved.

Proposition 1 and the lemma imply the following proposition.

{\bf Proposition 2.}
{\em In the framework of the conditions of Proposition 1, 
 the following bounds on the mass and scalar charge are  valid for all
$\mu >0$:
\bear{i5.16p}
   \frac{1}{2}\sqrt{h_{min} (Q_1^2 + Q_2^2)}  < GM, \\ \label{i5.16sp}
    |Q_{\varphi}| < |\lambda|_{max} \sqrt{h_{min} (Q_1^2 + Q_2^2)},
  \ear
for $\varepsilon = +1$ $(0< h_s < 2)$, and
\bear{i5.16m}
 \sqrt{\frac{1}{2} (Q_1^2 + Q_2^2)} < GM,
 \\ \label{i5.16sm}
  |Q_{\varphi}| < |\lambda|_{max} \sqrt{h_{max} (Q_1^2 + Q_2^2)},
  \ear
for $\varepsilon = -1$ $(h_s > 2)$}.

In proving  (\ref{i5.16sp}) and  (\ref{i5.16sm})  the following (obvious) relation was used:
$$h (- \mu + \sqrt{h^{-1} Q^2+ \mu^2}) = \frac{Q^2}{\mu + \sqrt{h^{-1} Q^2+ \mu^2}}.$$

In ref. \cite{ABDI} Propositions 1 and 2 were proved  for the case $\lambda_1 = \lambda_2$ ($h_1 = h_2$).
In this case the bound (\ref{i5.16p}) is coinciding (up to notations) with  the bound (6.16) from ref. \cite{GKLTT} (BPS-like inequality), which was proved there  by using certain  spinor techniques.

{\bf Remark 3.} {\em When one of  $h_s$, say $h_1$, is negative, 
the  conjecture is not valid. This may be verified just by analyzing the solutions with small enough charge $Q_2$.}

We note that here we were dealing with a  special class of solutions with phantom scalar field ($\varepsilon = -1$). Even in the  limiting case  $Q_2 = +0$ and $Q_1 \neq 0$ there exist  phantom black hole solutions which are not covered by our analysis \cite{CFR} (see also \cite{ACFR}.)

{\bf Remark 4.} {\em The inequalities on the mass (\ref{i5.13p}) and (\ref{i5.13m}) in Proposition 1 can be refined when $\lambda_1 \lambda_2 < 0$. For both cases
which are considered in Proposition 1, 
we get (see right panels of Figs. 1 and 2) 
\beq{i5.13pn}
   f(\mu,h_{min};Q^2) < GM <  f(\mu,h_{max};Q^2),
\eeq
where    $Q^2 = Q_1^2 + Q_2^2$ and 
$f(\mu,h;Q^2)$ is defined in (\ref{i5.18}).
The  bounds on mass (\ref{i5.13pn}) are a
specific feature of the model with two different dilatonic couplings of opposite sign. For $\lambda_1 \lambda_2 >0$, e.g. for  $\lambda_1 = \lambda_2$, one should use relations (\ref{i5.13p}) and (\ref{i5.13m}).
We also note that in the proof of Proposition 1 the condition $\lambda_1 \neq - \lambda_2$ was used. For the case $\lambda_1 = - \lambda_2$ the arcs on the right panels of Figs. 1, 2 reduce to points and we get $GM = f(\mu,h_{1};Q^2)$. }

%%%%%%%%%%%%%%%%%%%n%%%%%%%%%%%%%%%%%%%%%%%%%%%%%%%%%%%%%%%%%%%%%
\section{Conclusions}
%%%%%%%%%%%%%%%%%%%%%%%%%%%%%%%%%%%%%%%%%%%%%%%%%%%%%%%%%%%%%%%%

In this paper a family of non-extremal black hole dyon-like
 solutions in a 4d gravitational model
with a scalar field and two Abelian vector fields
 is presented. The scalar field is
either ordinary ($\varepsilon = +1$) or phantom  ($\varepsilon = -1$).
The model contains two dilatonic coupling constants
$\lambda_s \neq 0$, $s =1,2$, obeying $\lambda_1 \neq - \lambda_2$.

The solutions are defined up to two moduli functions $H_1(R)$ and
$H_2(R)$,  which obey two differential equations  of  second order
with boundary conditions imposed. For $\varepsilon = +1$ these
equations are integrable for four cases, corresponding
to the Lie algebras $A_1 + A_1$, $A_2$, $B_2 = C_2$ and $G_2$. In the first case ($A_1 + A_1$)
we have  $\lambda_1 \lambda_2 = 1/2$, while in the second one ($A_2$)
 we get $\lambda_1 = \lambda_2 = \lambda$ and  $\lambda^2 = 3/2$. Two other solutions, corresponding
 to the Lie algebras  $B_2 = C_2$ and $G_2$,  will be considered in separate publications.

There is also a special solution with dependent
electric and magnetic charges: $\lambda_1 Q_1^2 = \lambda_2 Q_2^2$, which is defined for all (admissible)  $\lambda_s$ and $\varepsilon$ obeying $\lambda_1 \lambda_2 > 0$.

Here we have also calculated some physical parameters of the solutions:
gravitational mass $M$, scalar charge $Q_{\varphi}$, Hawking temperature,
black hole area entropy and post-Newtonian parameters  $\beta$,  $\gamma$.
The PPN parameters $\gamma =1$ and $\beta$ do not depend upon
$\lambda_s$ and $\varepsilon$, if the values of $M$ and $Q_{\varphi}$ are fixed.

We have also obtained a formula, which relates  $M$,  $Q_{\varphi}$,
the dyon charges $Q_1$, $Q_2$, and the extremality parameter $\mu$
 for all values of  $\lambda_s \neq 0$. Remarkably, this formula does not contain $\lambda_s$
 and coincides with that of ref. \cite{ABDI}.
As in the case $\lambda_1 = \lambda_2$, the product of the Hawking temperature and
the Bekenstein-Hawking entropy do not depend upon $\varepsilon$,  $\lambda_s$
and  the moduli functions $H_s(R)$.

 Here we have obtained lower bounds on the gravitational mass and upper bounds on the scalar charge for
 $1 +2 \lambda^2_s \varepsilon > 0$,  which are based on the conjecture (from Sect. 5)
 on the parameters of solutions $P_1 = P_1(Q_1^2,Q_2^2)$,  $P_2 = P_2(Q_1^2,Q_2^2)$.
 In \cite{ABDI} we have  presented several  results of numerical calculations
 which support our bounds for $\lambda_1 = \lambda_2$.  
 A rigorous proof of this  conjecture may be the subject of a separate publication.
 For $\varepsilon  = +1$   the lower bound on the gravitational  mass is in agreement for $\lambda_1 = \lambda_2$ with that obtained earlier by Gibbons et al. \cite{GKLTT} by using certain  spinor techniques.
 
 It was noted in Sect. 3.3 that for  $\lambda_1 \neq \lambda_2$ there exist two integrable  cases corresponding to the Lie algebras $C_2$ and $G_2$, which will be analyzed in  separate papers. They do not occur for  
 $\lambda_1 = \lambda_2$.
  
An open question here is to find the conditions on the dilatonic coupling constants $\lambda_s$
which guarantee the existence of the second (hidden) horizon and the existence of the extremal black hole
in the  limit $\mu = +0$. For  $\varepsilon  = +1$, $\lambda_1 = \lambda_2$ this problem
was analyzed in refs. \cite{PTW,GKO}.  This question can be addressed to a separate publication.

\vspace{5pt}

{\bf Acknowledgment}

The authors acknowledge the support from the Program of target financing 
of the Ministry of Education and Science of the Republic of Kazakhstan Grant No. F.0755.
The paper was also funded by the Ministry of Education and Science of the Russian Federation
in the Program to increase the competitiveness of   Peoples' Friendship University (RUDN University) among the world's leading research and education centers in the 2016-2020 and  by the  Russian Foundation for Basic Research,  Grant  Nr. 16-02-00602.

\end{document}